\documentclass[journal=jacsat,manuscript=article]{achemso}

\usepackage{chemformula} 
\usepackage[T1]{fontenc} 
\usepackage[version=3]{mhchem} 
\usepackage{natbib}
\usepackage{graphicx}
\usepackage{bm}
\usepackage{color,soul}
\usepackage{hyperref}
\usepackage{float}
\usepackage{color,soul}
\usepackage{algorithm2e}
\hypersetup{colorlinks=true, urlcolor=blue, citecolor=blue}
\usepackage{threeparttable}
\usepackage{mathtools}
\SectionNumbersOn
\setcitestyle{square}
\author{Shashi B. Mishra$^{1}$}
\author{Somnath C. Roy$^{2}$}
\author{B. R. K. Nanda$^{1}$}
\email{nandab@iitm.ac.in}
\affiliation{$^{1}$Condensed Matter Theory and Computational Lab, Department of Physics,\\ Indian Institute of Technology Madras, Chennai- 600036, India \\
$^{2}$ Environmental Nanotechnology Lab, Department of Physics, \\Indian Institute of Technology Madras, Chennai- 600036, India}
\date{\today}

\title{Electronic Structure of Graphene/TiO$_2$ Interface: Design and Functional Perspectives}
  
\keywords{Interface electronic structure, bandgap, charge transfer, dipolar field, strain}
\begin{document}
\begin{abstract}
We propose the design of low strained and energetically favourable mono and bilayer graphene overlayer on anatase TiO$_2$ (001) surface and examined the electronic structure of the interface with the aid of first principle calculations. In the absence of hybridization between surface TiO$_2$ and graphene states, dipolar fluctuations govern the minor charge transfer across the interface. As a result, both the substrate and the overlayer retain their pristine electronic structure. The interface with the monolayer graphene retains its gapless linear band dispersion irrespective of the induced epitaxial strain. The potential gradient opens up a few meV bandgap in the case of Bernal stacking and strengthens the interpenetration of the Dirac cones in the case of hexagonal stacking of the bilayer graphene. The difference between the macroscopic average potential of the TiO$_2$ and graphene layer(s) in the heterostructure lies in the range 3 to 3.13 eV, which is very close to the TiO$_2$ bandgap ($\sim$ 3.2 eV). Therefore, the proposed heterostructure will exhibit enhanced photo-induced charge transfer and the graphene component will serve as a visible light sensitizer. 
\end{abstract}

\section{\label{sec:level1}Introduction}
Owing to superior carrier mobility, large surface area, high strength and unique electronic properties, graphene has been extensively investigated for fabricating nanocomposites with semiconducting materials such as TiO$_2$, ZnO and SnO$_2$, etc.\cite{Kamat2010,Zhang2010,Zhang2011,Xiang2012a,Morales-Torres2012,Cao2017,Low2017,Liu2018} The resultant heterostructures show improved efficiency in photocatalysis and photovoltaic applications due to enhanced charge carrier separation and shift in the absorbance spectra to visible region.\cite{Manga2009,Wang2012,JiangDu2011,Lee2012,Xiang2012,Kiarii2018,Zhang2017} Among the semiconductors, anatase TiO$_2$ is still the most preferred material for photocatalytic and photovoltaic applications due to its chemical stability, increased photo-corrosion resistance and suitable electronic band edge positions.\cite{Hashimoto2005,Linsebigler1995,Henderson2011,Zheng2020} The study of electronic structure of the graphene-TiO$_2$ (G/TiO$_2$) interface therefore holds significance from both the fundamental and the applied perspectives. Several reports have discussed G/TiO$_2$ interface which we summarized in Table~\ref{tab1} and the relevant issues are further discussed.

Functionalization of monolayer or bilayer graphene is also equally significant from the application point of view.\cite{Xie2013,Singh2015,Shashi2020}. For example, inducing a bandgap in this zero bandgap semiconductor has remained an open issue.\cite{Hu2017,Du2011a} In the present context, the opening of bandgap is important for the effective absorption of light, which may help in photocatalytic/photovoltaic applications. In fact, there are a number of first principle electronic structure calculations that have reported graphene as an overlayer on anatase (A) TiO$_2$(001) surface leading to a finite bandgap of 0.45$-$0.65 eV.\cite{Gao2013,Yang2017,Yang2013a} Interestingly, similar calculations indicate that the graphene does not show any bandgap when it becomes an overlayer on other TiO$_2$ surfaces such as A(101)\cite{Li2013,Ferrighi2016,MARTINS2018} and rutile (R) (110).\cite{Du2011,Long2012,Gillespie2017} Therefore it raises two valid possibilities. Firstly, the surface of A(001) may induce a large epitaxial strain on the graphene layer by means of an effective $\pi$-band model. Theoretically, it is reported that a bandgap appears when a monolayer graphene is uniaxially strained along the zigzag direction with a strain above 26 \%\cite{Pereira2009} or with the application of a shear strain above 16 \%.\cite{Choi2010} It is also reported that anisotropic strain, combined with a $-$20 \% compression in the armchair direction and 11 \% stretching in the zigzag direction, may produce a bandgap in graphene.\cite{Kerszberg2015} Secondly, there may be strong chemical bonding between graphene and TiO$_2$ which leads to breakdown of the characteristics of Dirac $\pi$-bands.

Furthermore, the charge transfer mechanism at the G/TiO$_2$ interface is still subject to intensive investigation due to its impact on potential applications in photocatalytic and photovoltaic processes.\cite{Low2017,Sandeep2017,Manga2009,Xie2013,MARTINS2018,Liu2018} An experimental photoluminescence (PL) spectroscopy study carried out on a multilayer of hybrid film made up of alternate graphene and titanium nanosheets reported that the transfer of photo-excited electrons from graphene to titanium layers is faster than the average excited state carrier lifetime for graphene with measured electron transfer time scales of the order of 200$-$250 fs.\cite{Manga2009} This observation is also theoretically supported, in the case of G/R(110) interface, by non-adiabatic MD (NAMD) simulations where the photo-excited electron transfer is reported to be several times faster than electron-phonon energy relaxation.\cite{Long2012} Similar results on charge transfer from C-$p$ to Ti-3$d$ states have also been reported for G/A(101)\cite{Li2013,Ferrighi2016,MARTINS2018} and G/R(110) interfaces.\cite{Du2011,Gillespie2017} Irrespective of the orientation of the surface, it is expected that similar kind of charge transfer should occur for G/A(001) interface as well. However, it is reported that the charge transfer takes place from A(001) surface to graphene.\cite{Gao2013,Yang2017} 

In this study, the electronic structure and stability of the G/A(001) interface is examined through density functional calculations and an energetically favorable low strained graphene overlayer is proposed. Analysis of the electronic structure reveals that, in the absence of chemical bonding, the heterostructure is stabilized through van der Waals interaction and, as a result, both the substrate and the overlayer retain their pristine electronic structure. The charge density redistribution arising from the heterointerface formation leads to the creation of an interfacial dipole. The dipolar fluctuation is also responsible for a minor charge transfer only across the interface and, in the case of bilayer graphene, creates a potential gradient between the upper and lower carbon layers. This results in the opening of a narrow bandgap of $\sim$ 80 meV in the case of AB-stacked bilayer graphene. The difference in the macroscopic potential across the graphene and TiO$_2$ is found to be close to the bandgap of TiO$_2$ which makes the graphene Dirac point to be in resonance with the TiO$_2$ conduction band. This facilitates the photo-induced charge transfer and makes the heterostructure promising for photocatalytic and solar cell applications. The earlier report of wide bandgap opening in G/A(001) interface\cite{Gao2013,Yang2013a,Yang2017} is reviewed and a detailed analysis of the band dispersion in the full Brillouin zone reveals that graphene retains its characteristic Dirac feature irrespective of the strain induced by the host TiO$_2$. Further, a comprehensive analysis of role of strain on the electronic structure of monolayer graphene is presented, which demonstrates that with uniaxial strain a bandgap can not open in monolayer graphene.

\begin{table}
\scriptsize
\begin{tabular}{c c c c c c c c c c c}
\hline \hline
Literature & system & TiO$_2$ & graphene & \multicolumn{2}{c} {strain (\%)} & no. of & $E_g$ & $E_{ad}$ & $d$ & $\Delta Q$ \\
&  &supercell  & supercell  & x  & y & atoms & (eV) & (eV/C) & (\AA) & (e)\\
 \hline
Gao et al.,2013\cite{Gao2013} & G/A(001)\textsuperscript{\emph{a}} & $4\times4$ &$7\times6\textsuperscript{\emph{a}}$ & - & - & 276 & 0.47 & -0.02 & 2.85 & A to G \\
Yang et al.,2013\cite{Yang2013a} & G/A(001)\textsuperscript{\emph{b}} & $21.98\times11.80$ & - & 4.04 & - &- &-& & 1.60 & C-O bond \\
  & G/A(101)\textsuperscript{\emph{b}} & $10.02\times21.98$ & - & 3.17 & - &- &1.32& &3.34& A to G\\
  & G/A(110)\textsuperscript{\emph{b}} & $10.36\times10.36$ & - & 5.31 & - &-&1.54&- &3.185& A to G\\
Yang et al.,2017\cite{Yang2017} & G/A(001) & $6\times2$ &$3\times10$ & 3 & - & 204 & 0.53 &-0.07 & 2.82 & A to G \\
Du et al.,2011\cite{Du2011} & G/R(110) & $5\times2$ &$6\times6$& & 0.02 & 251 & 0 & -0.02 & 2.75 & G to R\\
Gillespie et al.,2013\cite{Gillespie2017} & G/R(110) & $2\times5$ &$3\times6$ & -0.30 & 2.32 & 152 & 0 & 0.02 & 2.90 & G to R\\
Li et al.,2013\cite{Li2013} & G/A(101) & $2\times2$ &$5\times3$ & 4.23 & 1.47 & 78 & &-0.05 & 2.51 & G to A \\
Ferrighi et al.,2016\cite{Ferrighi2016} & G/A(101) & $2\times2$ & $5\times3$ & 2.15 & -3.45 & 72 & 0 & -0.03 & 2.97 & G to A \\
\hline \hline
\caption{\label{tab1}Summary of reported structures, and related parameters (such as size of the supercell, strain, number of atoms, bandgap ($E_g$), interfacial adhesion energy ($E_{ad}$), layer separation ($d$) and charge transfer ($\Delta Q$)) for several G/TiO$_2$ (R and A) interfaces.}
\end{tabular}
\textsuperscript{\emph{a}} The initial construction of $7 \times 6$ rectangular graphene unit cell breaks the periodic boundary condition. As a result of this the graphene layer undergoes asymmetric rotation which is discussed later in the present work.\\
\textsuperscript{\emph{b}} The literature presented the supercell in \AA{} unit rather than in terms of unit cells.
\end{table}

\section{Methodology and Computational Details}
\textit{Ab initio} electronic structure calculations are performed using pseudopotential approximations and PBE-GGA exchange-correlation functional as implemented in Quantum Espresso.\cite{Giannozzi2017} Ion-electron interactions are expressed through ultrasoft pseudopotential. The electron wave functions are expanded using plane-wave basis set with a kinetic energy cutoff of 30 Ry and an augmented charge density cutoff of 300 Ry. Dispersion corrections have been included through the semi-empirical Grimme-D2 van der Waals correction.\cite{grimme2006} The calculated lattice parameters of anatase in the bulk phase are a = 3.794 \AA, c = 9.754 \AA{} which agree well with the previously reported experimental and theoretical values.\cite{Burdett1987,Mo1995,Mishra2018} Anatase (001) surface is created using slab geometry with four TiO$_2$ layers (thickness $\sim$ 8.5\AA), which is observed to be sufficiently thick enough\cite{Gao2013,mishra2020} and a 15 \AA{} vacuum along the out-of-plane direction. The optimizations are performed using a $4\times4\times1$ Monkhorst-Pack k-mesh with force convergence criteria set to be 0.025 eV/\AA{}. During relaxation, only the upper two layers are allowed to move freely, while the atoms in the lower two layers are fixed to their bulk positions. For the calculation of the electronic structure, the Brillouin zone integration is performed on a finer grid of $8\times8\times1$ $k$-mesh. In order to calculate the electronic structure of the pristine and strained graphene unit cells, a denser $k$-mesh of $60\times60\times1$ is considered as it is found that the Dirac point is very sensitive to the strain and requires a finer $k$-mesh. In order to achieve an accurate prediction of band-crossing at the G/A(001) interface, an appropriate $k$-mesh around the energy minima is estimated and energy diagonalization is performed at those points from which the $k$-path is determined and taken into account in the calculation of the band structure. DFT+$U$ calculations have been carried out on the composite structure to examine if there is a correlation effect on the electronic structure. Details of the results are given in the Supplementary Information.  

The average strain induced on the graphene overlayer due to the lattice mismatch with the surface TiO$_2$ is estimated as\cite{Stradi2017}
 \begin{equation}\label{eq:1} 
    \Bar{\epsilon}=\frac{\mathopen|\epsilon_{xx}\mathclose|+\mathopen|\epsilon_{yy}\mathclose|+\mathopen|\epsilon_{xy}\mathclose|}{3},
 \end{equation}
 where $\epsilon_{xx}$, $\epsilon_{yy}$ and $\epsilon_{xy}$ are the components of the 2D strain tensor. These components are expressed through primitive surface lattice vectors ($a_{1,x}$, $a_{1,y}$) and ($a_{2,x}$ and $a_{2,y}$) of TiO$_2$ and primitive lattice vectors of graphene ($b_{1,x}$, $b_{1,y}$), and ($b_{2,x}$, $b_{2,y}$) as follows:
  \begin{eqnarray}\label{eq:2} 
     \epsilon_{xx} &=& \frac{a_{1,x} - b_{1,x}}{a_{1,x}}, \nonumber 
     \epsilon_{yy} = \frac{a_{1,y} - b_{1,y}}{a_{1,y}}, \nonumber \\
     \epsilon_{xy} &=& \frac{1}{2}\frac{b_{2,x}a_{1,x} - b_{1,x}a_{2,x}}{a_{1,x}a_{2,y}}
 \end{eqnarray}

The binding between the graphene layer and A(001) surface is quantitatively measured through the  adhesion energy given by the following expression.
 \begin{eqnarray} 
    E_{ad} &=& E_{G/A(001)} - E_{A(001)}^{opt} - E_{G}^{strained},\label{eq:3}\\
    E_{strain} &=& E_{G}^{strained} - E_{G}^{ideal}. \label{eq:4} 
 \end{eqnarray}
Where, $E_{G/A(001)}$ and $E_{A(001)}^{opt}$ are the total energies of the optimized G/A(001) composite and pristine A(001) surface, respectively. $E_{G}^{strained}$ and $E_{G}^{ideal}$ represent the total energies of strained graphene layer and ideal graphene supercell ($7\times3$), respectively. $E_{strain}$ is the energy cost to strain the graphene layer to align on the TiO$_2$(001) surface. The deformation energy is estimated using the following equation~\cite{Gillespie2017}.
 \begin{eqnarray}\label{eq:5} 
     E_{def} &=& (E_{A(001)}^{def} - E_{A(001)}^{opt}) + (E_{G}^{def} - E_G^{strained}).
 \end{eqnarray}
Where, $E_{A(001)}^{def}$ and $E_G^{def}$ represent the total energies of  pristine A(001) surface and strained graphene layer as in the composite structure. Therefore, the binding energy of the system becomes 
\begin{eqnarray}\label{eq:6} 
    E_{BE} &=& E_{ad} - E_{def} - E_{strain}.
\end{eqnarray}
The other way to analyse the interaction between graphene and A(001) surface is through the charge redistribution that takes place across the interface. It is measured through the three-dimensional charge density difference of G/A(001) composite estimated by the following formula.
   \begin{equation}\label{eq:7} 
    \Delta \rho(r) = \rho_{G/A(001)}(r) - \rho_{A(001)}(r)- \rho_G(r)
   \end{equation}
where, $\rho_{G/A(001)}$ represents the charge density of G/A(001) heterostructure; while $\rho_{A(001)}$ and $\rho_G$ represent the charge densities of pristine A(001) surface and graphene layer, respectively in the same coordinate space as that of the heterostructure G/A(001).

\section{Results and discussions}
\subsection{\label{design}Design of G/A(001) interface}
\begin{figure}[hbt!]
\centering
 \includegraphics[width=15cm,height=10.2cm]{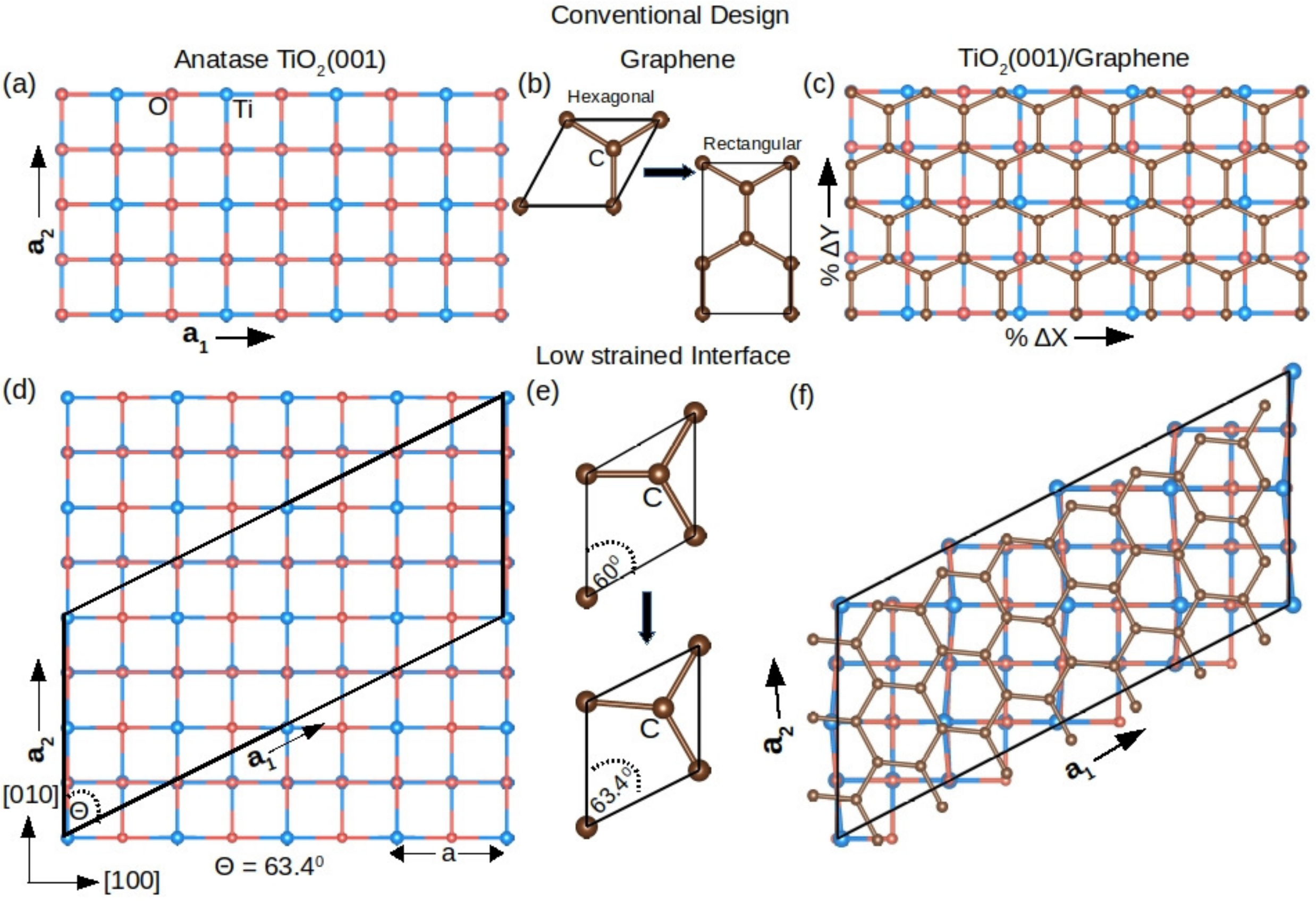}
 \caption{\label{model-design}Construction of the interface between A(001) TiO$_2$ surface and graphene. (a) A 4$\times$2 A(001) slab with in-plane lattice translation vectors ($\vec{a_1}$ and $\vec{a_2}$); (b) transformation of graphene unit cell from hexagonal to rectangular cell; (c) the composite structure G/A(001) interface for which the strain values on graphene are +2.74 \% and -12.28 \% along $\hat{x}$ and $\hat{y}$-directions, respectively. The rectangular unitcell of graphene is used in the previous works\cite{Gao2013,Yang2013a} or in the design of G/R(110) interface.\cite{Du2011,Gillespie2017} (d) The lattice vectors for A(001) surface for which angle is 63.43$^{\circ}$ with a supercell size of $2\sqrt{5}\times2$ having 24 Ti and 48 O atoms in a four layered TiO$_2$ slab. (e) The rotation of the graphene unit cell from 60$^{\circ}$ to 63.4$^{\circ}$. (f) The optimized structure of graphene layer with a supercell size of $7\times3$ on the A(001) surface. The average strain in this case is calculated to be $\sim$ 1.50\%.}
\end{figure}

The key to design a low-strain G/A(001) interface lies in patterning the graphene layer on TiO$_2$(001) surface as shown in Fig.~\ref{model-design}. In the conventional method, as followed in earlier works,\cite{Gao2013,Yang2013a} a rectangular unit cell is created from the regular hexagonal graphene unit as shown in Fig.~\ref{model-design}(b). A supercell of desired size to match the A(001) surface is then created from this unit cell and placed over the TiO$_2$ surface to create a commensurate structure (Fig.~\ref{model-design}c). However, the lattice matching needs a strain to be applied along the zigzag and armchair directions of the graphene layer. The magnitude of the strain depends on the size of the surface area. Similar formalism has been considered for the G/R(110) interface.\cite{Du2011,Gillespie2017}

In this section, we present a scheme to define various low strained interfaces by designing new corresponding unit cells for the graphene layer, the vectors of which are given as:
  \begin{equation}\label{eq:8}
    \begin{pmatrix} \vec{a_1} \\ \vec{a_2} \end{pmatrix} 
    =\begin{pmatrix} $ma$ & $na$\\ 0 & 2a \end{pmatrix}
    \begin{pmatrix} \hat{x} \\ \hat{y} \end{pmatrix} 
\end{equation}
where, a is the lattice parameter of A(001) slab in the $xy$-plane, m and n are integers, $\Theta$ = $\angle \vec{a_1}\vec{a_2}$. Depending on the values of m and n, the angle between the lattice vectors $\vec{a_1}$ and $\vec{a_2}$ of A(001) surface varies. Few selected (m, n) pairs and corresponding $\Theta$ are listed in Table~\ref{tab2}. The table also lists the size of commensurate TiO$_2$ surface unit cells, size of the graphene supercell, total number of atoms in the composite considering a four layers of TiO$_2$ slab, the lattice mismatch, average strain as obtained from Eq.~\ref{eq:1}. The vector $\vec{a_2}$ is fixed with a length of 2a (= 7.588 \AA) as that accommodates almost three primitive unitcells of graphene (3a$_{G}$ = 7.378 \AA) and the lattice mismatch is $\sim$ 2.74 \%. For the case of $\Theta$ = 90\textdegree{}, we show that by simply increasing the surface area, one can reduce the average strain as it gives the flexibility to find out an ideal supercell of the rectangular graphene unit so that a matching area of the graphene can be obtained. However, despite there is a good matching among the supercell, the local strain at the unitcell range will persist. Furthermore, large surface area leads to large number of atoms in the composite structural unit, which in turn makes it computationally very expensive.

\begin{table*}[h]
\scriptsize
\centering
 \caption{\label{tab2}Unit cell construction: the commensurate unit cell between the A(001) surface and the graphene layer along with the strain tensor ($\epsilon$). The negative and positive signs in $\epsilon$ represent the compression and elongation of the graphene. The terms $\epsilon_{xx}$, $\epsilon_{xy}$ and $\epsilon_{yy}$ represent the components of $\epsilon$~\cite{Stradi2017} and $\bar{\epsilon}$ is average strain as defined in Eq.~\ref{eq:1}. Here, we have considered four layered TiO$_2$ slab.}
\begin{tabular}{ c c c c c c c c c c c}
  \hline \hline
  m & n & Angle & A(001) & Graphene &  Total no. & \multicolumn{4}{c}{Strain (\%)} & $E_{ad}$\\
  & & $\Theta$($^{\circ}$)& supercell & supercell & of atoms & $\epsilon_{xx}$ & $\epsilon_{yy}$ & $\epsilon_{xy}$ & ($\bar{\epsilon}$) & (eV/C atom)\\
  \hline
 4 & 0 & 90 & $4\times2$ & $6\times2$\textsuperscript{\emph{a}} & 144 & +2.74 & -12.28 & 0.0 & 5.01 & -0.036 \\
 4 & 0 & 90 & $4\times8$ & $6\times7$\textsuperscript{\emph{a}} & 552 & +2.74 & +0.77 & 0.0 & 1.17 & \\
 6 & 0 & 90 & $6\times6$ & $9\times5$\textsuperscript{\emph{a}} & 612 & +2.74 & +6.43 & 0.0 & 3.05 & \\
 5 & 2 & 68 & $\sqrt{29}\textsuperscript{\emph{*}}\times2$ & $7\times3$\textsuperscript{\emph{b}} & 168 & +10.17 & +2.74 & 0.0 & 4.30 & -0.031\\
 4 & 2 & 63.4 & $2\sqrt{5}{\emph{$^\dagger$}}\times2$ & $7\times3$\textsuperscript{\emph{b}} & 138 & +1.75 & +2.74 & 0.0 & 1.50 & -0.040\\
  \hline \hline
 \end{tabular}\\
 $\sqrt{29}$\textsuperscript{\emph{*}}  = (5a$_{1,x}$, 2a$_{1,y}$); $2\sqrt{5}$\textsuperscript{\emph{$^\dagger$}}  = (4a$_{1,x}$, 2a$_{1,y}$).\\
 \textsuperscript{\emph{a}}{rectangular graphene unit cell with lattice parameters $\vec{b_1}$ = 2.46 \AA{} and $\vec{b_2}$ =
 4.26 \AA.}\\
 \textsuperscript{\emph{b}}{hexagonal graphene unit cell with lattice parameter $\vec{b}$ = 2.46 \AA{}.}
\end{table*}

We looked at several possible values of $\Theta$ and found that the case of $\Theta$= 63.4\textdegree{} represents an average lattice mismatch of 1.50\%. The other designs are presented in Fig.~\ref{models-other-angle}. Also, the composite unit with $\Theta$ = 63.4\textdegree{} forms a natural construct for graphene, as it is close to angle between the primitive lattice vectors of the pristine graphene. The resulting G/A(001) interface is shown in Fig.~\ref{model-design}(f). In addition to the applied strain, stability plays a key role in forming the interface. In the case of heterostructures, the stability is estimated by the calculation of the adhesion energy as given in the Eq.~\ref{eq:3}. The adhesion energy for the designed interface is calculated to be -1.49 eV ($\sim$ -0.04 eV/C atom), while a strain ($E_{strain}$) of $\sim$ 4.33 eV (0.10 eV/C atom) is needed to form the graphene overlayer from the ideal graphene $7\times3$ supercell (Eq.~\ref{eq:4}). The deformation energy is estimated to be $\sim$ 0.03 eV (0.001 eV/C atom). This is a combination of TiO$_2$ deformation energy (0.02 eV) and strained graphene deformation energy (0.01 eV) (see Eq.~\ref{eq:5}). This amounts to a binding energy ($E_{BE}$) of -5.86 eV ($\sim$ -0.14 eV/C atom) as estimated using Eq.~\ref{eq:6}.

\subsection{\label{dos}Electronic structure of the low strained G/A(001) interface}
\begin{figure}[hbt!]
\centering
\includegraphics[width=16cm,height=8cm]{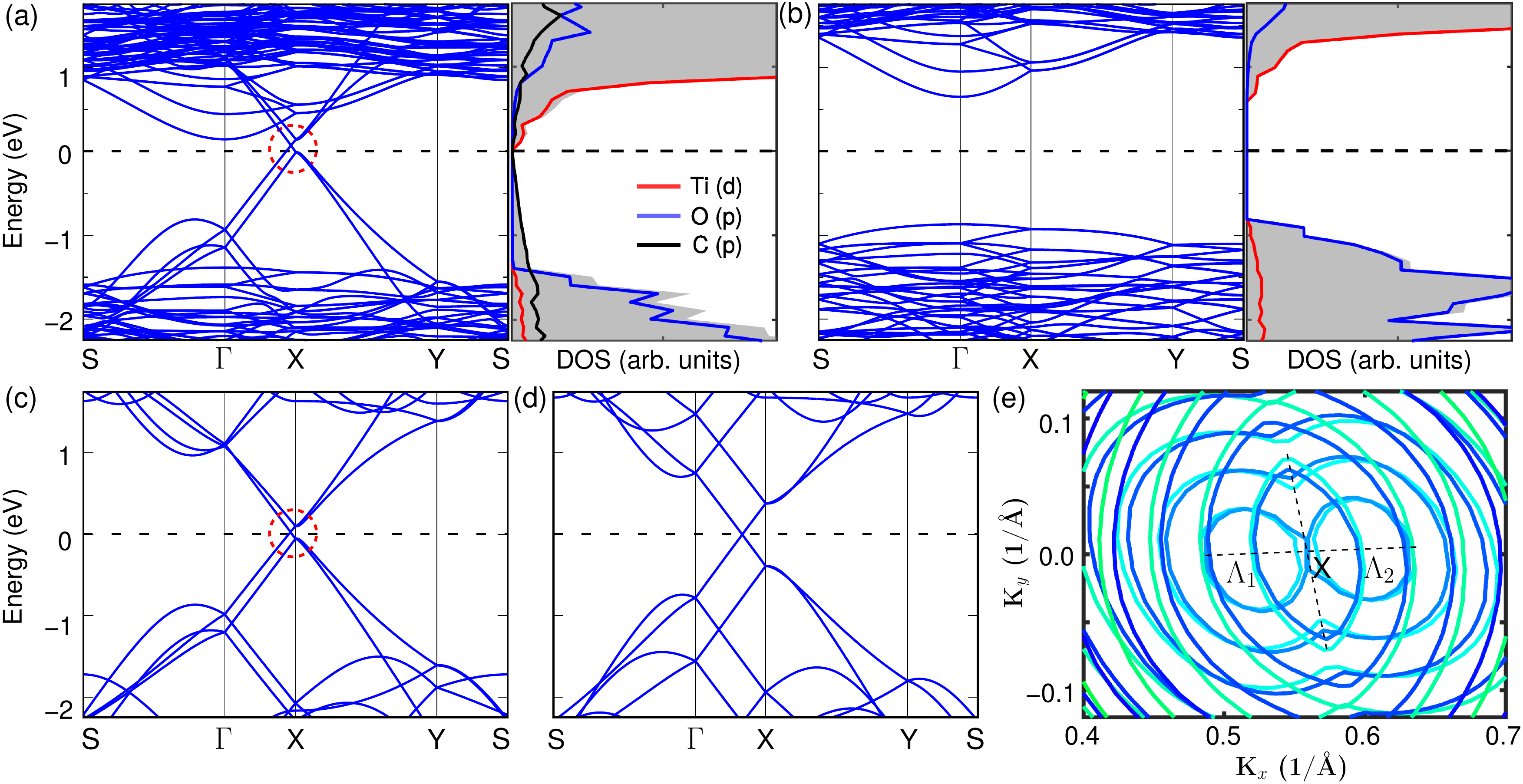}
\caption{\label{band-dos}(a) The band structure and total DOS along with the partial DOS of Ti-$d$, O-$p$ and C-$p$ orbitals at the G/A(001) interface and (b) the isolated A(001) slab. The total DOS is shown in gray shaded area. (c) The band structure of the isolated graphene layer same as that of the G/A(001) heterostructure and (d) the band structure of the pristine $7\times3$ graphene layer. (e) The contour indicates a pair of low-energy Dirac cones at $\Lambda_1$ and $\Lambda_2$, while the other directions represent high-energy states. In this figure, the blue and green lines correspond to valence and conduction bands, respectively.}
\end{figure}
The electronic structure of the low-strained G/A(001) interface is examined by calculating the band dispersion, the densities of states (DOS), the charge transfer across the interface and the average potential profiles. The band structure and the DOS are plotted in Fig.~\ref{band-dos}. The band structure of the heterostructure (Fig.~\ref{band-dos}a) is found to be almost a superposition of the band structures of TiO$_2$ slab (Fig.~\ref{band-dos}b) and the isolated strained graphene layer (Fig.~\ref{band-dos}c). It is further substantiated from the total and partial DOS. The Ti-$d$ and O-$p$ DOS reveal the retention of the semiconducting behaviour of the TiO$_2$, while C-$p$ states occupy the Fermi level. As GGA underestimates the bandgap, often GGA+$U$ calculations are carried out to match the bandgap. To examine if correlation has impact on the states of the heterostructure at the Fermi level, we have carried out GGA+$U$ calculations and the results are provided in the Supplementary Information. As expected only the hybridized Ti-$d$ and O-$p$ are affected by the $U$ creating a wider bandgap, while the states in the vicinity of the Fermi surface remain unchanged (see Fig.~\ref{surface-dos-u7}). We therefore draw the following two important conclusions. First, the dispersion of the Dirac band in graphene layer is unperturbed due to the formation of the interface. Second, the van der Waals interactions emerging from the fluctuating dipoles govern charge redistribution at the interface. In order to examine the effect of strain on the graphene, we compared its band structure with that of the pristine graphene calculated using a $7\times3$ supercell (see Fig.~\ref{band-dos}d). The down-folding of the bands with the shrink in the Brillouin Zone leads to a shift in the Dirac point and now it lies along the $\Gamma-$X path. The strain, which is compressive along one of the lattice vectors and expansive along the other, brings the Dirac point closer to X. This is likely due to strengthening of the $p_z-p_z$ bond in one direction and the weakening along the other. Figure~\ref{band-dos}(e) shows the iso-contours of the top valence band and bottom conduction band on the $k_x-k_y$ plane. We find that at two points, $\Lambda_1$ and $\Lambda_2$, the Dirac valleys are formed. 

\begin{figure}[hbt!]
    \centering
    \includegraphics[width=6cm,height=6.5cm]{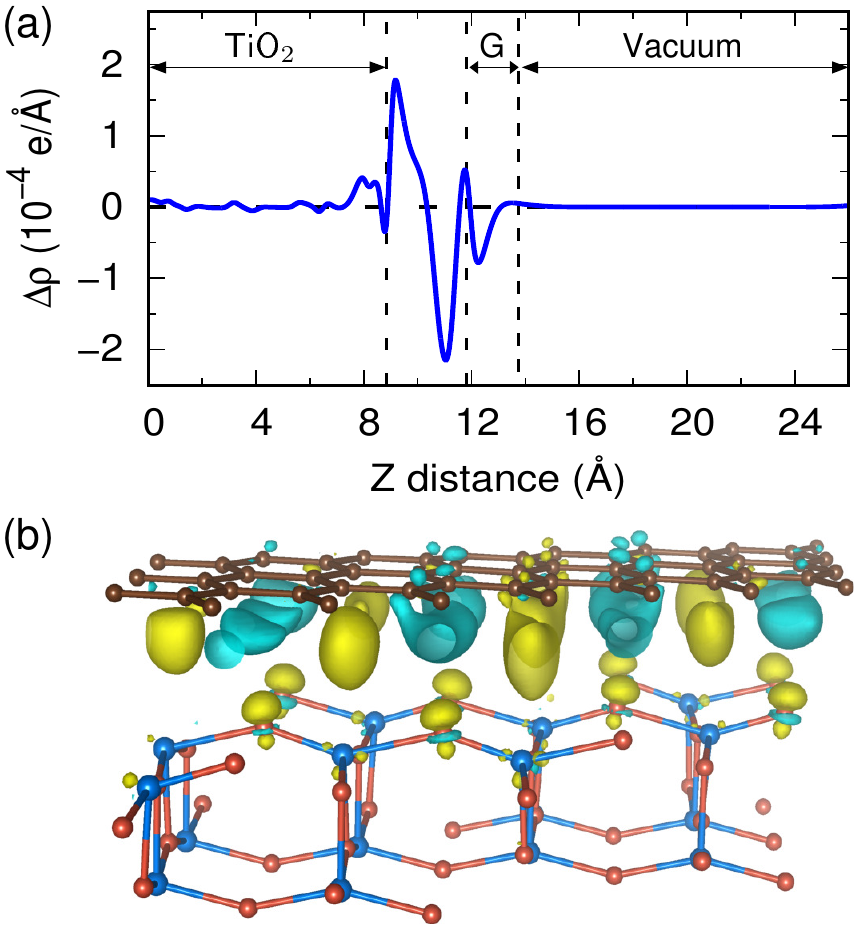}
    \caption{\label{chargediff}(a) The planar-averaged differential charge density $\Delta \rho (z)$ for the G/A(001) heterostructure as a function of position in the z-direction. The vertical line indicates the location of the top layer of the A(001) and the graphene layer. (b) Charge density difference for the heterostructure calculated using Eq.~\ref{eq:7}. The yellow and cyan colors represent the charge accumulation and depletion regions, respectively. The iso-value is set to be $6\times10^{-4}$ e/\AA$^3$.}
\end{figure}

The heterostructure phenomena is further investigated by calculating the charge transfer across the interface as shown in Fig.~\ref{chargediff}(a). Here, the positive and negative values correspond to charge accumulation and depletion regions, respectively. It shows that the electrons are mainly accumulated on the top layer of the TiO$_2$ surface, while graphene layer acts as an electron donor. Figure~\ref{chargediff}(b) presents planar-averaged charge density difference along the slab direction $Z$. The observations made from this figure are as follows. First, the inner layers of the TiO$_2$ slab hardly participate in the interaction and the redistribution of the charge takes place at the G/A(001) interfacial region. Second, it shows that electrons are transferred from the graphene layer to the TiO$_2$(001) surface, which is opposite to what was reported earlier on G/A(001).\cite{Gao2013,Yang2013a,Yang2017} But this observation is consistent with previous G/A(101)\cite{Ferrighi2016} and G/R(110)\cite{Long2012} interfaces. Third, the amount of the charge transfer is small (of the order 10$^{-4}$ e/\AA$^3$) which further confirms the lack of ionic or chemical bonding and only a weak van der Waals coupling stabilizes the G and A(001) surface. To quantify the change in charge on each of the atoms in the G/TiO$_2$ heterostructure as a result of dipolar fluctuation, we have performed Bader charge analysis on the heterostructure and compared the charges with their pristine components as presented in Table S1. The net average charge on the C-atom is estimated to be $4 \times 10^{-4}$ e which is very small, while the net charge on the TiO$_2$(001) surface is $-2 \times 10^{-4}$ e as compared to their neutral pristine components. To get a clear picture of this interfacial charge transfer, we plotted the three-dimensional charge density difference calculated using Eq.~\ref{eq:7} in Fig.~\ref{chargediff}(b). The yellow and cyan color lobes represent the charge accumulation and depletion regions, respectively. The charge transfer at the interface is primarily through van der Waals interaction and to verify the presence or absence of covalent interaction, we have plotted the Crystal Orbital Hamiltonian Population (COHP) for a pair of closest C and O atom shown in Fig.~\ref{cohp-tio2}, which as expected is negligible across the energy window suggesting the absence of covalent bonding.

\subsection{Electronic Structure of the Reported Composites: The missing Dirac Point}
\begin{figure}[hbt!]
\centering
\includegraphics[width=16.5cm, height=7cm]{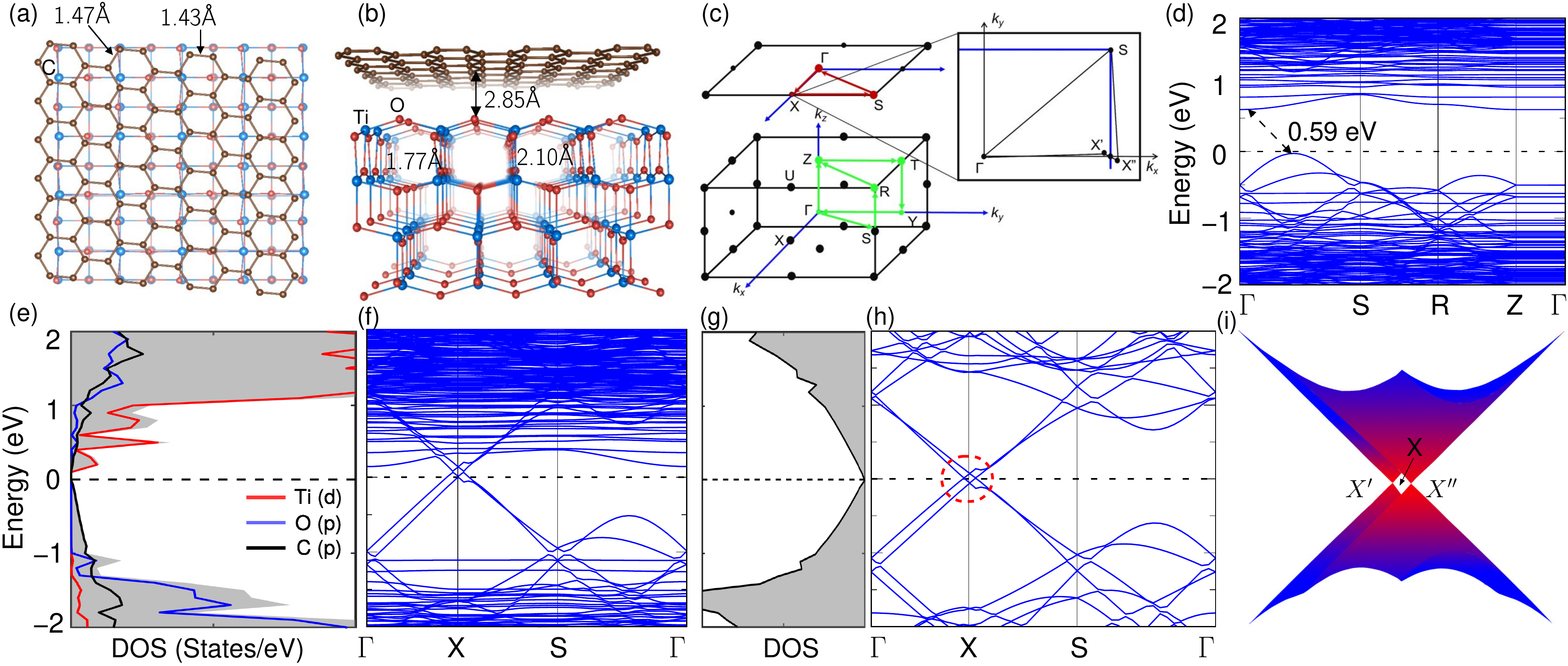}
\caption{\label{gao_band}(a) Top view and (b) side view of the relaxed structure of the G/A(001) heterostructure as described in literature~\cite{Gao2013} using a rectangular graphene unit cell. (c) The bulk BZ and its surface mapping are marked with high symmetry points. As the slabs are considered in this study, the bulk BZ is virtual. Shifting of the Dirac point from the high symmetry point 'X' is shown in the inset for clarity. The bulk BZ $k$-path is shown in green which is exactly similar to the path taken in the previous study.\cite{Gao2013,Yang2013a,Yang2017} The band structure is reproduced in (d). (e) The total (gray shaded) and partial DOS (color) and (f) the band structure calculated along the surface $k$-path. (g,h) The total DOS and the band structure for the isolated graphene layer (same graphene layer as in the G/A(001) heterostructure, but with A(001) slab removed) respectively. (i) The 3D-band structure of the isolated graphene layer in the vicinity of the Dirac point. The Dirac cone is formed at two points $X^\prime$ and $X^{\prime\prime}$.}
\end{figure}

While we proposed a low strained interface where no bandgap appears, earlier literature reported a bandgap as large as 0.59 eV in the G/A(001) interface.\cite{Gao2013,Yang2013a,Yang2017} It is therefore prudent to examine whether a bandgap can be opened in the G/A(001) interface. For this purpose, we have taken the example of the interface proposed in an earlier report~\cite{Gao2013} which is designed out of a supercell of size $14.95\times14.87\times26.13$ \AA$^3$. In terms of unit cells, it is formed out of a $7\times6$ graphene layer and a four layered thick $4\times4$ A(001) surface. The structure has been further relaxed. However, to be consistent with the reported configuration, the bottom two layers are kept fixed during the structural relaxation and the graphene layer is maintained at a distance of 2.85 \AA{} from the top of the A(001) surface. Figure~\ref{gao_band}(a) and (b) show the optimized G/A(001) interface structure. The graphene layer is strained as the C$-$C bond length increases to 1.47 and 1.44 \AA{} along the zigzag and armchair directions, respectively. In TiO$_2$, the equatorial Ti$-$O bond is elongated on one side (2.11 \AA) and compressed on the other (1.76 \AA) compared to the bulk value of 1.79 \AA. The Brillouin zone (BZ) of the corresponding structure is shown in Fig.~\ref{gao_band}(c). The bulk BZ is artificial as the periodicity along (001) includes the slab plus the vacuum periodicity with the mapping of bulk to surface BZ, and the high symmetry (HS) points in the reciprocal space for the G/A(001) are marked.

In Fig.~\ref{gao_band}(d), we have shown the band structure along the $k$-path (see the green lines in Fig.~\ref{gao_band}(c)) used in the earlier study~\cite{Gao2013} and the band dispersion is very well reproduced, giving a band gap of $\sim$ 0.59 eV. However, when the system is strained anisotropically, the geometry of the direct lattice space and thus the reciprocal space changes and, in such cases, the paths connecting the high symmetry points do not necessarily reveal the salient features of the band structure. It is therefore necessary to scan the eigenstates in the entire BZ and, hence, examining the DOS is a correct approach. In Fig.~\ref{gao_band}(e), we have shown the total DOS and it yields a zero bandgap. The calculation of the partial DOS indicates that in the vicinity of the Fermi level, while the C-$p$ states constitute the valence bands, both the C-$p$ and Ti-$d$ states form the conduction band.

A concrete understanding of the band structure requires an examination of the top of the valence and the bottom of the conduction band and of the orbital composition. Specifically, it is crucial to examine whether the graphene has lost its Dirac bands and whether the bands are now changed due to covalent interaction and/or the charge transfer with the host TiO$_2$ layer. To address this, we scanned the points around the high symmetry point X. This is because, if we consider the rectangular unit cell for graphene as shown in Fig.~\ref{model-design}(b) instead of the original hexagonal unit cell (Fig.~\ref{model-design}e), the Dirac bands touch each other at X. Figure~\ref{gao_band}(f) indeed shows that there are two pairs of linear bands that cross each other at X$^{\prime}$ and X$^{\prime\prime}$ close to X and resemble that of a pristine graphene. In addition, there are a bunch of nearly non-dispersive bands below and above the Fermi level separated by a large gap. Further to identify the bands predominated by the C-$p$ states, we constructed a hypothetical structure where we retained the strained graphene layer and removed the TiO$_2$ slab of the composite. The resulting total DOS and band structure are shown in Fig.~\ref{gao_band}(g) and (h). It is almost identical to the heterostructure band structure, ignoring the non-dispersive bands. This indicates that there is no covalent bonding between the host TiO$_2$ and the graphene overlayer. In spite of a large strain, the latter retains its Dirac cone as can be seen from the band structure plotted on the $k_x$-$k_y$ plane around X. Later, we explained why strain alone can not open a bandgap in the G/A(001) interface when the graphene is monolayer.

\subsection{Bilayer graphene on A(001) surface}
\begin{figure}[hbt!]
\centering
\includegraphics[width=16cm,height=8.5cm]{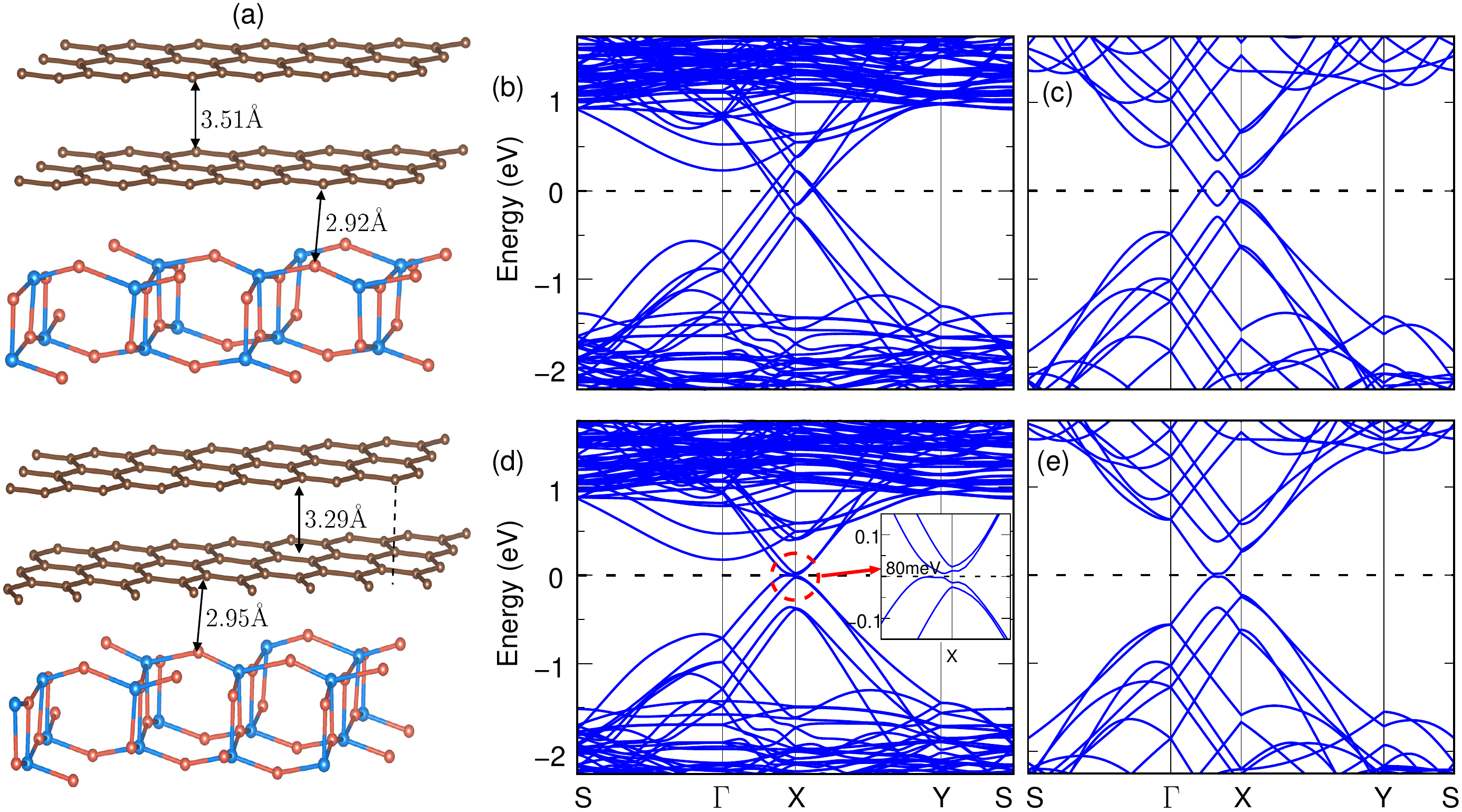}
 \caption{\label{bilayer-band}(a) The optimized geometry of bilayer graphene overlayer in both hexagonal (AA) and Bernal (AB) stacking on TiO$_2$(001) surface. (b) The band structure of AA-stacked A(001) heterostructure, and (c) the band structure of pristine $7\times3$ supercell of AA-stacked bilayer. Similarly, the electronic structure for (d) AB-stacked A(001) heterostructure and (e) pristine $7\times3$ supercell of AB-stacked bilayer graphene. The inset shows the zoomed portion of the AB-stacked overlayer on A(001) surface.}
\end{figure}

It is often easy to experimentally synthesize multilayer graphene that are equally promising as monolayer graphene for applications.\cite{Liu2018} We will therefore examine the interfacial electronic structure of bilayer graphene (BLG), both with Bernal (AB) and hexagonal (AA) stacking, grown on the A(001) surface. The optimized structures and the resulting electronic structures are shown in Fig.~\ref{bilayer-band}. In the case of AB (AA) stacking, the separation between the A(001) surface and the BLG is found to be 2.95 \AA{} (2.92 \AA{}), while the separation between the carbon layers is measured as 3.29 \AA{} (3.51 \AA{}) which is almost the same as in the case of a freestanding BLG.\cite{Nanda2009} The bilayer overlayer is found to be stable and the adhesion energies are calculated to be -0.0780 and -0.0776 eV per unit cell respectively for AB and AA stacking.

To understand the interaction between the bilayer graphene and the TiO$_2$ surface, we plotted the band structure of the heterostructures and compare it to that of the pristine bilayer supercell as shown in Fig.~\ref{bilayer-band}(b-e). In the case of pristine system, the AA-stacking shows interpenetrating Dirac cones (Fig.~\ref{bilayer-band}c) and the AB-stacking results in zero bandgap with parabolic conduction and valence bands touching each other at the Dirac point (Fig.~\ref{bilayer-band}e).~\cite{Nanda2009} In the case of AA-stacked BLG, the interfacial electronic structure is simply a combination of the electronic structure of TiO$_2$(001) surface and graphene similar to that of monolayer overlayer. However, in the case of AB-stacked BLG, a minor gap appears as a result of dipole fluctuation discussed in the following paragraph. The GGA+$U$ band structure for the bilayer overlayer, which is presented in Fig.~\ref{bilayer-u7}, also reveals a similar band structure except shifting of Ti-$d$ dominated conduction and valence bands. To substantiate further the effect of interface, the band structure of freestanding strained bilayer graphene is shown in Fig.~\ref{bilayer-strained} and it infers that strain alone does not induce any bandgap.

\begin{figure}[hbt!]
    \centering
    \includegraphics[width=6.5cm,height=7.5cm]{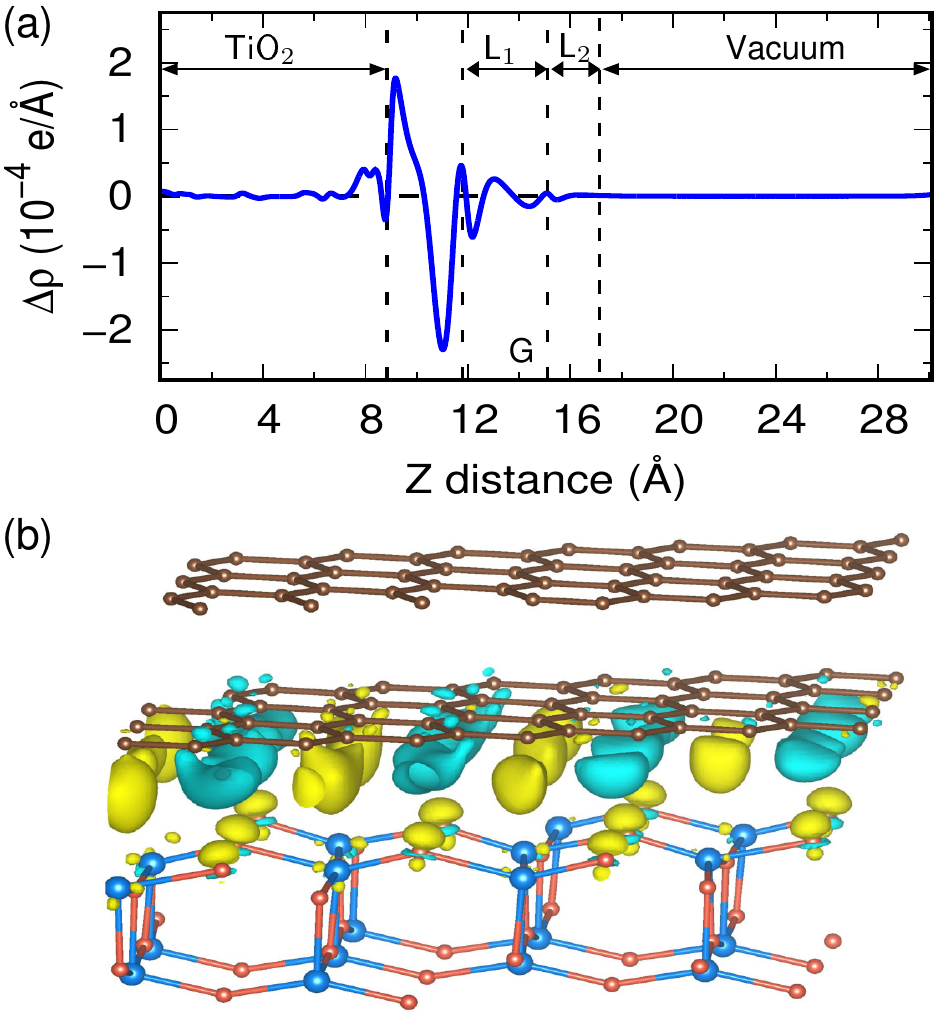}
    \caption{\label{ABchargediff}(a) The planar-averaged differential charge density $\Delta \rho (z)$ for the AB-stacked BLG on the TiO$_2$(001) surface as function of $z$-coordinate, and (b) the charge density difference plot. The yellow and cyan colors represent the charge accumulation and depletion regions, respectively. The iso-value is set to be $6\times10^{-4}$ e/\AA$^3$.}
\end{figure} 

In Fig.~\ref{ABchargediff}, we estimated the charge transfer across the interface in the case of AB stacking, and found that a weak charge transfer, purely governed by the dipole fluctuations, occurs only between the bottom graphene layer and the surface layer of TiO$_2$. The induced dipolar field breaks the symmetry between the two graphene layers by creating a potential difference.~\cite{Ohta2006} Both the strain induced by the substrate and the potential difference affect the band structure in the vicinity of the Fermi surface as can be seen from the inset of Fig.~\ref{bilayer-band}(d). There are two pairs of parabolic bands on the Fermi surface. The small hump in the conduction band and the dip in the valence band at the high-symmetry point X is the result of the dipolar field. The field leads to the formation of a very small bandgap ($\sim$ 80 meV). The charge transfer in the case of AA-stacked BLG is almost identical to the case of AB-stacking, and is therefore not shown here. However, since the electric field does not affect the linear dispersion of the bands in the AA-stacked BLG, the interpenetration of the Dirac cones remains unaffected, as can be seen from Fig.~\ref{bilayer-band}(b).
\begin{figure}[hbt!]
    \centering
    \includegraphics[width=16cm,height=4.8cm]{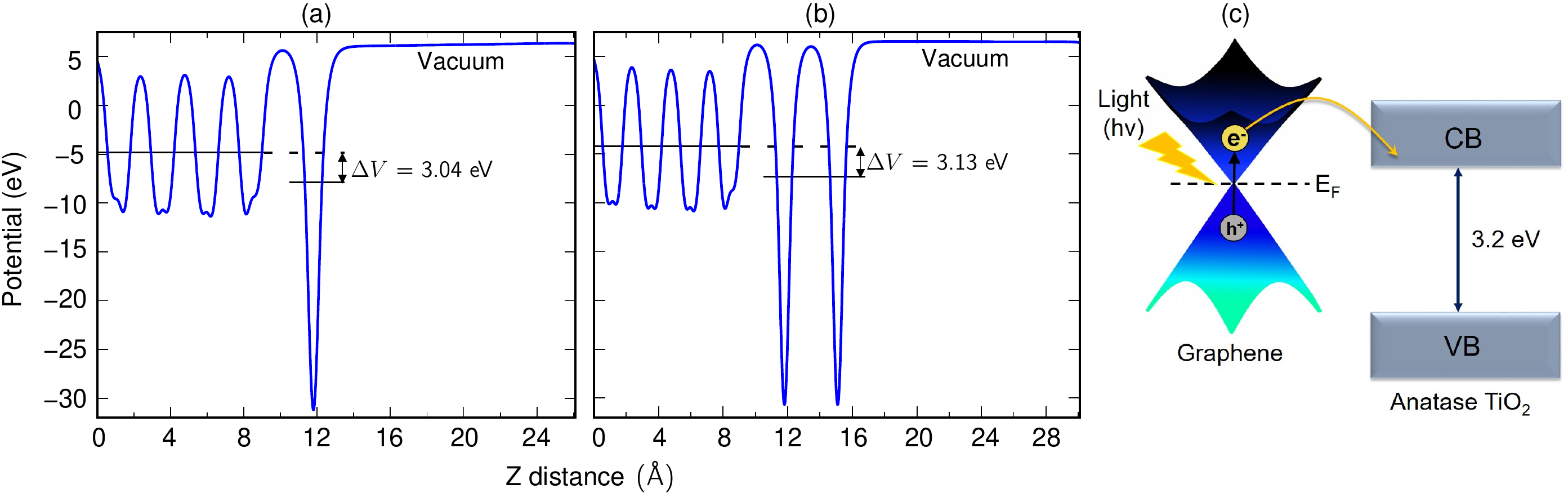}
    \caption{\label{avg_pot}The planar average (blue lines) and macroscopic average (black lines) of the electrostatic potential for the G/A(001) heterostructure with (a) monolayer and (b) AB-stacked bilayer graphene overlayer along the normal $z$-direction. (c) A schematic illustration of the charge transfer process at the G/A(001) surface using visible light. The band edge position of the G and A(001) surface is predicted by alignment of their DOS with respect to the vacuum level (see Fig.~\ref{dos-align}).}
\end{figure} 

The heterojunction potential leading to the band-offset between the graphene and the TiO$_2$ surface is examined by estimating the macroscopic average of the electrostatic potential as shown in Fig.~\ref{avg_pot}(a) and (b). The surface of TiO$_2$ has a higher electrostatic potential than that of the monolayer and bilayer graphene. The potential difference is calculated to be 3.04 eV and 3.13 eV respectively for mono and bilayer graphene. The schematic of the charge transfer process at the G/TiO$_2$ interface is presented in Fig.~\ref{avg_pot}(c), in which the band edge positions are drawn by aligning their DOS with respect to vacuum level (see Fig.~\ref{dos-align}). Since the G/A(001) interface is a zero bandgap system, it can absorb all wavelengths of light. When visible light incidents, electrons are excited in the graphene layer. These excited electrons are in resonance with the TiO$_2$ conduction band and can move easily from C-$p$ states to the Ti-$d$ states in the conduction band. In this way, graphene acts as a visible light sensitizer as well as charge separator. Since, the bandgap at the Fermi surface can be easily tuned through applied electric field. In the case of AB-stacking, the TiO$_2$/AB heterostructure represents a better platform for gas sensing, photo-degradation and other optical absorption studies.

\subsection{Effect of Strain}
In the absence of a chemical bonding between the substrate and the graphene overlayer, the root question, which we now need to address, is whether the monolayer graphene bandgap can be opened by strain alone? The theoretical model proposed earlier suggests that a bandgap can be induced with a strain above $\sim$ 26\% along the zigzag direction\cite{Pereira2009} or with a shear strain of $\sim$ 16\%.\cite{Cocco2010} However, the model was limited to the $\pi$-bands, as it has generally been assumed that the electronic structure of graphene is governed by the $\pi$-band and that the planar $\sigma$-bands have largely contributed to structural stability. In one the report, it has been found that defects and disorder can bring the $\sigma$ states close to the Fermi surface.\cite{Choi2010} The strain can be applied both along the armchair and zigzag directions. However, it has been conclusively shown that the strain along the armchair direction does not open a gap.\cite{Pereira2009,Choi2010} The electronic structure of the graphene with the strain applied along the zigzag direction will therefore be discussed in this section.
\begin{figure}[hbt!]
 \centering
\includegraphics[width=16.5cm,height=5cm]{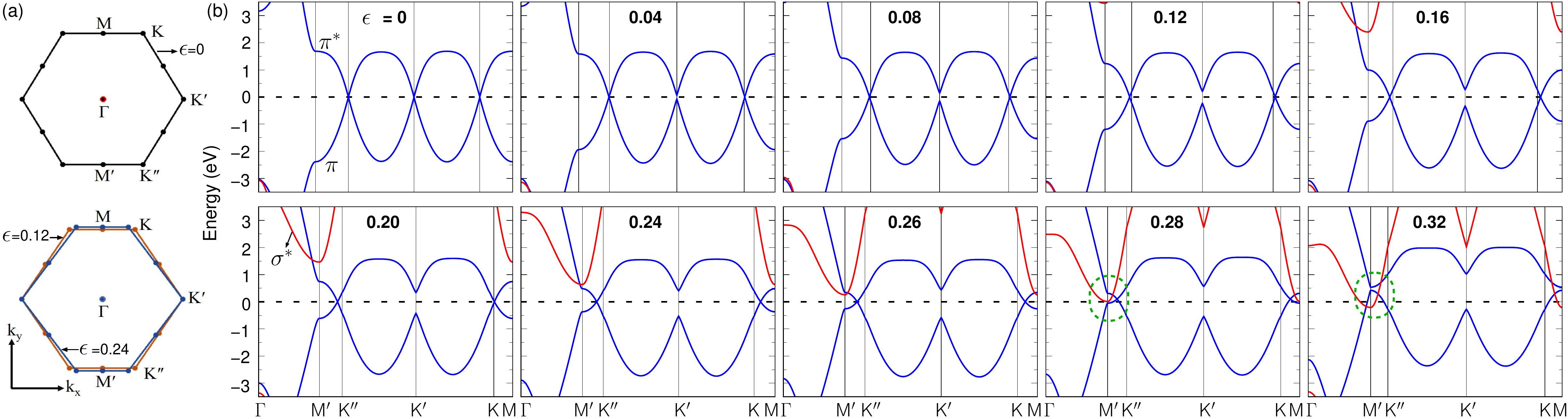}
 \caption{\label{dft-band}(a) The Brilloun zone of graphene for unstrained (upper one) and strained (bottom one) cases ($\epsilon$ = 0.12 and 0.24). High symmetry points are indicated. (b) The evolution of the monolayer graphene band structure with the strain $\epsilon$ along the zigzag direction. The $\pi$ and $\pi^{*}$ bands are shown in blue and the lower lying $\sigma^{*}$ band is shown in red. The Fermi level is set to zero.} 
\end{figure}

As a result of the applied strain, the lattice parameters are modified as \textbf{a$_{i}^\prime$} = \textbf{I}(1+$\epsilon$)$\cdot$\textbf{a$_i$}, where \textbf{I} is the $2\times2$ unitary matrix, \textbf{a$_i$} the unstrained lattice vector, and $\epsilon$ is the strain parameter.\cite{Gomez-Arias2016} The associated reciprocal lattice vectors \textbf{b$_i$} are modified as \textbf{b$_i^\prime$} = $(\textbf{I} + \epsilon)^{-1} \cdot$\textbf{b$_i$}. The evolution of the band structure with the increasing strain ($\epsilon$) is shown in Fig.~\ref{dft-band}. With strain, the linear dispersion of the $\pi$-bands does not change. However, the point of crossing (Dirac point) of these bands gradually shifts. With strain, K$^\prime$ and K$^{\prime \prime}$ gradually move to M and M$^\prime$, and as a consequence, K$^{\prime}$ no longer remains as a Dirac point as a gap opens up at this point. Dirac points earlier located at K and K$^{\prime\prime}$ are now moved to M and M$^\prime$. Beyond $\epsilon$ = 0.28, a narrow bandgap opens between $\pi$ and $\pi^*$ bands. The lower $\sigma^*$ band (red curve) that was far above the Fermi level is gradually pushed down with strain and above $\epsilon$ = 0.28, it crosses the Fermi level at M and M$^{\prime}$, and therefore, there is no real bandgap in the system.

\section{Conclusion}
To summarize and conclude, the interfacial electronic structure for the anatase TiO$_2$-graphene heterostructure grown along the (001) direction is examined, and a low-strained and energetically favorable orientation for the graphene overlayer is proposed.  Graphene almost retains the natural angle of $\pi/3$ between its two lattice vectors in this design. In the absence of covalent bonding and ionic charge transfer, the system is stabilized by van der Waals interactions. In the case of monolayer graphene, the electronic structure of the heterostructure is found to be a combination of the pristine TiO$_2$ and graphene electronic structure. In contrast to previous reports, the monolayer graphene (MLG) does not open a gap and the linear dispersion of the Dirac cones is maintained. This was attributed to the fact that, in the absence of covalent bonding, the epitaxial strain induced by TiO$_2$ could not open a bandgap in the MLG. The bilayer graphene as overlayer reconstructs the Fermi surface. In this case, as dipolar fluctuation led charge transfer exists only between the surface of TiO$_2$ and the bottom layer of the BLG, an electrostatic potential gradient is induced between the graphene layers. The potential gradient are capable of creating a bandgap ($\sim$ 80 meV) in the AB-stacked BLG overlayer. Both MLG and BLG provide excellent platforms for photo-induced charge transfer from graphene to TiO$_2$ as the average potential difference between them is in the range of 3.0 to 3.13 eV, which is close to the TiO$_2$ bandgap ($\sim$ 3.2 eV), therefore the G/TiO$_2$ heterostructure is promising for photocatalytic and solar cell applications.

\begin{acknowledgement}
The authors acknowledge the support from Defense Research and Development Organization, India, through Grant No. ERIP/ER/RIC/201701009/M/01 and HPCE, IIT Madras for providing computational facilities. SBM would like to thank M. Gupta for useful discussions.
\end{acknowledgement}
\appendix \addtocontents{toc}{\protect\setcounter{tocdepth}{0}}

\section{\label{other-model}Structural and Electronic Analysis of other G/A(001) model Interfaces}
\begin{figure}[hbt!]
\includegraphics[width=16cm,height=7.5cm]{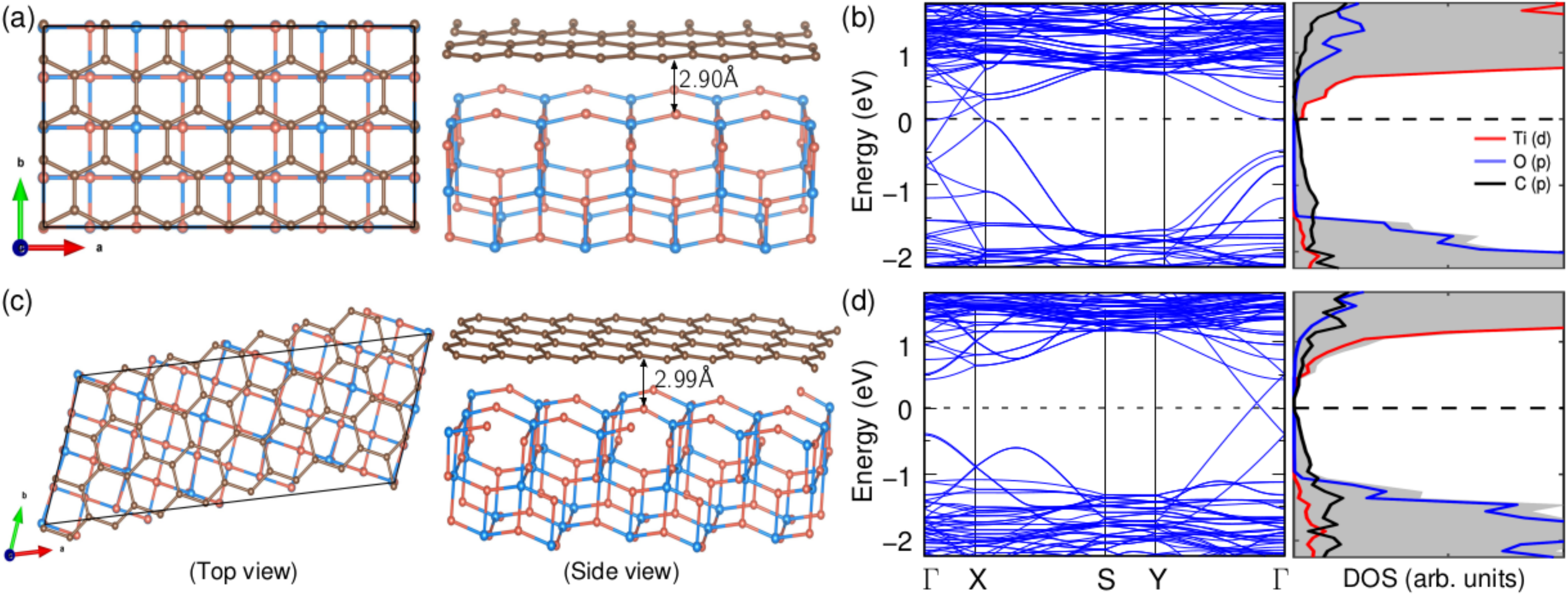}
 \caption{\label{models-other-angle}Construction of G/A(001) interface as a function of graphene overlayer at different angles  which are chosen from the angle made by the in-plane lattice translation vectors (\textbf{a$_1$} and \textbf{a$_2$}) of A(001) surface. (a) Top view and side view of the G/A(001) interface at angle ($\Theta$) 90\textdegree{} and (b) its corresponding band structure and partial DOS along with the total DOS. Similarly, (c) shows the optimized structure at $\Theta$ = 68.19\textdegree{} and (d) its band and partial DOS. The total DOS is shown in gray shaded region.} 
\end{figure}

As indicated in the section~\ref{design} by aligning the graphene overlayer at different angles (such as 90\textdegree{} and 68.19\textdegree{}) governed through m and n values of Eq.~\ref{eq:8}, new commensurate unit cells are constructed as shown in Fig.~\ref{models-other-angle}. For $\Theta$ = 90\textdegree{}, the overlayer design is the same as that of the rectangular pattern of graphene as shown in Fig.~\ref{model-design}(a-c). The separation between the TiO$_2$ slab and graphene for this case is calculated to be 2.90 \AA{} (Fig.~\ref{models-other-angle} a,b). The average value of strain calculated using Eq.~\ref{eq:1} and the $E_{ad}$ using Eq.~\ref{eq:3} for this interface is estimated to be 5.01 \% and -0.04 eV/C respectively. Similarly, for $\Theta$ = 68.19\textdegree{} the graphene overlayer remains at a distance of $\sim$ 3 \AA{} from the top of TiO$_2$ slab (Fig.~\ref{models-other-angle} c and d). The interfacial average strain is estimated to be $\sim$ 4.30\% and the $E_{ad}$ of -0.03 eV/C. The electronic structure and PDOS for both the overlayer patterns ($\Theta$ = 90\textdegree{} and 68.19\textdegree{}) show that the graphene retains its Dirac cone nature (Fig.~\ref{models-other-angle} b and d) and there is no bandgap in system. The repositioning of the band is due to the different strain values applied on the graphene overlayer. The analysis follows as that discussed for $\Theta$ = 64.43\textdegree{} in Sec.~\ref{dos}.

\section{Electronic structure of G/TiO$_2$(001) with $U$ = 7 eV}
\begin{figure}[hbt!]
\centering
\includegraphics[width=16cm,height=4.2cm]{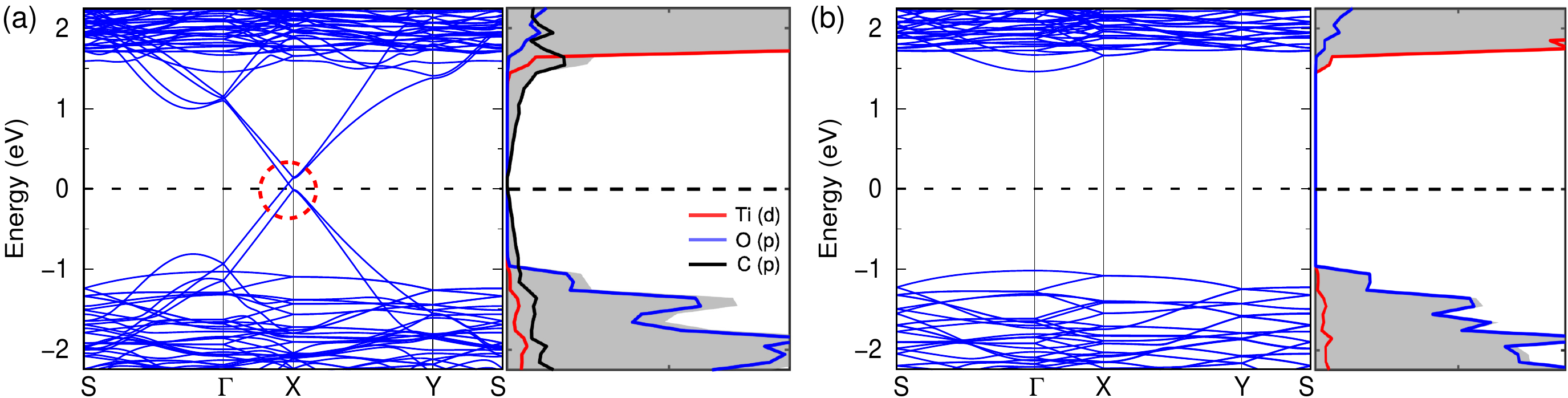}
 \caption{\label{surface-dos-u7} (a) The band structure and partial DOS of Ti-$d$, O-$p$ and C-$p$ states along with the total DOS for low strained G/TiO$_2$(001) interface with $U$ = 7 eV. (b) The band and DOS for the TiO$_2$ (001) surface with $U$ = 7 eV.}
\end{figure}

With $U$ = 7 eV, the electronic band structure and partial DOS of Ti-$d$, O-$p$ and C-$p$ states are plotted as shown in Fig.~\ref{surface-dos-u7}(a). As compared to the PBE-GGA results presented in Fig.~\ref{band-dos}(a) with finite $U$ (= 7 eV), the Ti-$d$ states in shifted away from the Fermi energy. The bands appearing at the Fermi energy are from the C-p states, and hence the gapless feature of the composite remains intact. The TiO$_2$ (001) surface band and DOS with $U$ = 7 eV are shown in Fig.~\ref{surface-dos-u7}(b). 

\section{Electronic structure of BLG/TiO$_2$(001) with $U$ = 7 eV}
\begin{figure}[hbt!]
\centering
\includegraphics[width=13cm,height=5cm]{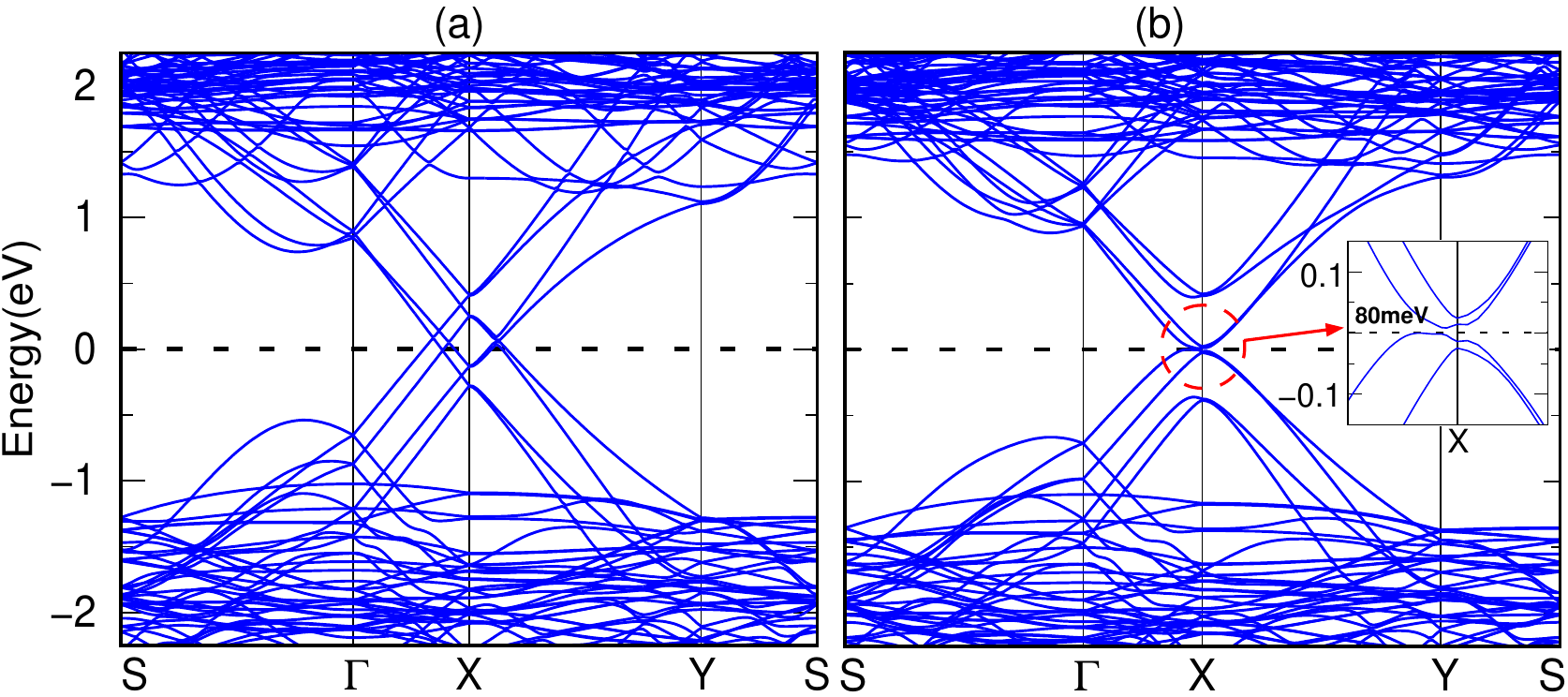}
 \caption{\label{bilayer-u7}The band structure of (a) AA- and (b) AB- stacked bilayer graphene layer over TiO$_2$ (001) surface with $U$ = 7 eV. It shows the band nature near the Fermi energy remains the same except the shifting of Ti-$d$ states above the Fermi level.}
\end{figure}

With $U$ = 7 eV, the band structure of AA and AB-stacked bilayer graphene overlayer on the TiO$_2$(001) surface are displayed in Fig.~\ref{bilayer-u7}. It shows similar band feature as compared to that in the absence of $U$, except shifting of Ti-$d$ states dominated in the conduction band. The AB-stacked bilayer opens up a minor band gap ($\sim$ 80 meV) due to potential gradient arising out of the weak charge transfer between the TiO$_2$(001) surface top layer and the closest carbon layer. 
\section{Electronic structure of strained bilayer graphene}
\begin{figure}[hbt!]
\centering
\includegraphics[width=13cm,height=5cm]{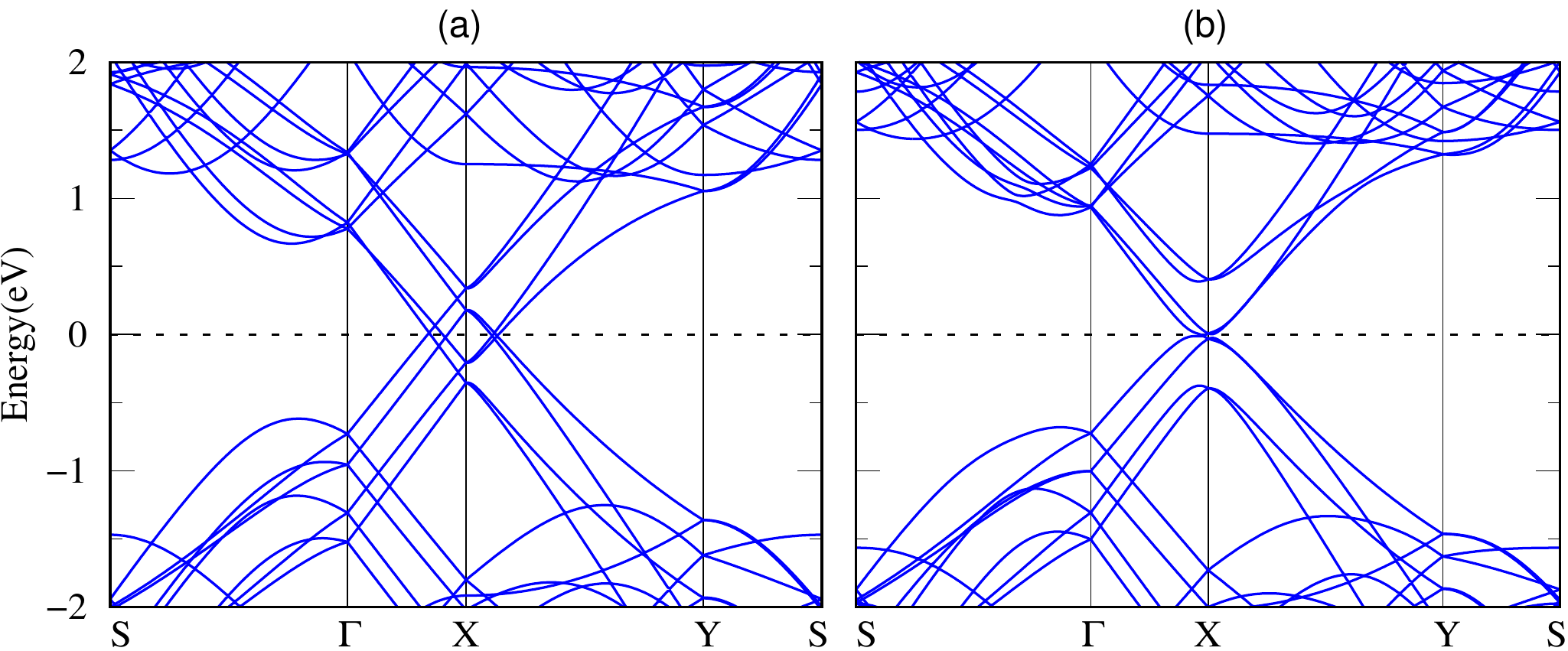}
 \caption{\label{bilayer-strained}The band structure of strained (a) AA-stacked and (b) AB-stacked bilayer graphene. It clearly shows there is no band gap opening in both the cases. The gap only opens up when there is potential gradient between the two layers in AB-stacked bilayer graphene.}
\end{figure}

The band structure of strained AA and AB-stacked bilayer graphene (BLG) is shown in Fig.~\ref{bilayer-strained}. It shows that there is no band gap opening in the BLG, for which it is reported that strain alone cannot open any band gap, but the electric field can open up a gap in the system and depending on the magnitude of the electric field the gap changes. 

\section{Alignment of DOS of Graphene and TiO$_2$(001) Surface}
\begin{figure}[hbt!]
\centering
\includegraphics[width=16cm,height=8cm]{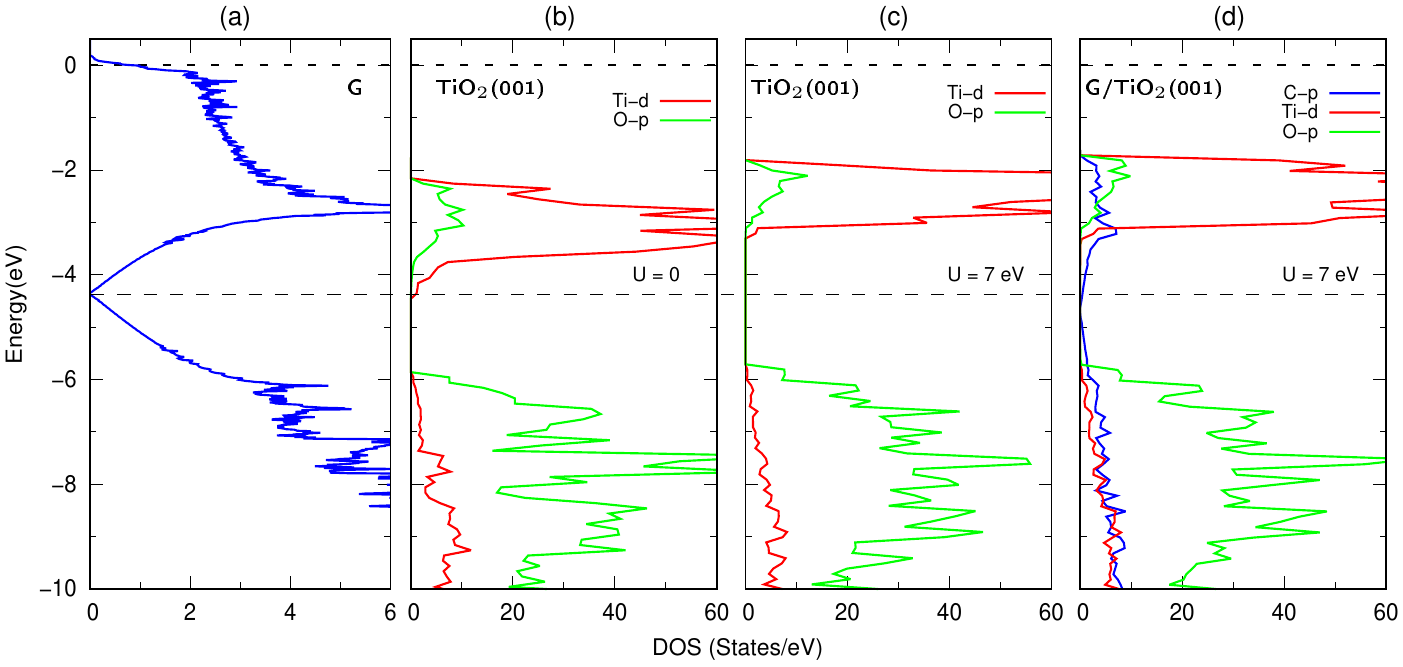}
 \caption{\label{dos-align} Density of states alignment of graphene Dirac crossing with the TiO$_2$ surface with respect to vacuum level. (a) Total density of states (DOS) of graphene layer. Partial DOS of Ti-$d$ and O-$p$ orbitals of the TiO$_2$ (001) surface (b) without any $U$ and (c) with $U$ = 7 eV. (d) The partial DOS of C-$p$, Ti-$d$ and O-$p$ orbitals G/TiO$_2$ heterostructure. The vacuum level is reference as zero energy.}
\end{figure}

To understand the position of graphene Dirac cone with respect to TiO$_2$ (001) surface, we have plotted the total density of states (DOS) of the graphene layer and the partial DOS of Ti-$d$ and O-$p$ states with respect to vacuum level as shown in the Fig.~\ref{dos-align}. It shows without any correlation correction, the conduction band formed by the Ti-$d$ states are at the same level as that of the graphene Dirac state (see Fig.~\ref{dos-align}(a,b)). With the application of $U$ = 7 eV, the Ti-$d$ states are pushed upwards in energy and hence the graphene states lie between the valence band formed by O-p states and the conduction band minima formed by the Ti-$d$ states (Fig.~\ref{dos-align}(c)).

\section{Bader Charge Analysis of G/TiO$_2$(001) interface}
To quantify the change in charge on each of the atoms in the G/TiO$_2$ heterostructure as a result of dipolar fluctuation, we have performed Bader charge analysis on the heterostructure and compared the charges with their pristine components. The net average charge on the C-atom is estimated to be $4\times10^{-4} e$ which is very weak, while the net charge on the TiO$_2$ (001) surface is $-2\times10^{-4} e$. To verify our charge analysis, we calculated the charge on the pristine graphene layer and TiO$_2$ (001) surface which correctly gives to zero. Hence, as a result of the heterostructure formation, there is a very weak charge transfer from the graphene to the TiO$_2$ surface which was also inferred from the planar average charge density plot as shown in Fig.~\ref{chargediff}(a).

\section{COHP analysis of a pair of closest C and O atoms}
\begin{figure}[hbt!]
\centering
\includegraphics[width=5.0cm,height=7.0cm]{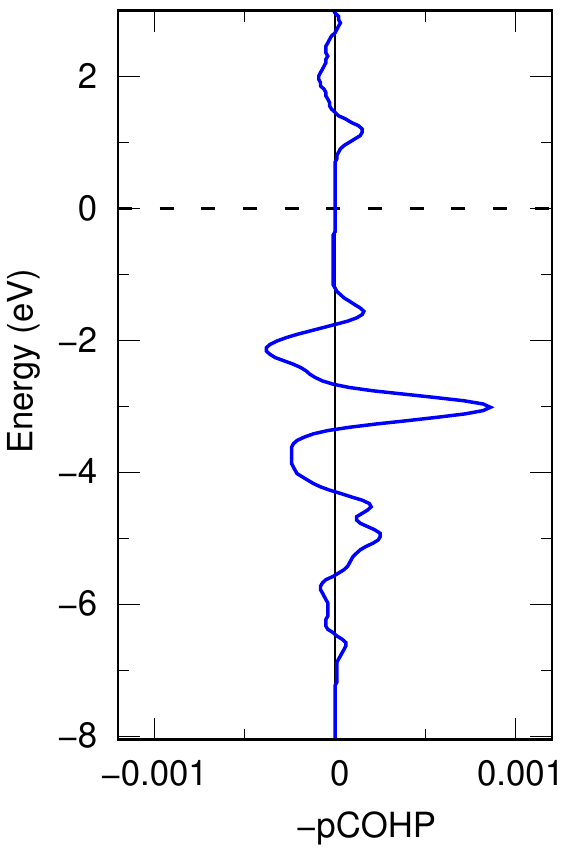}
 \caption{\label{cohp-tio2} The $-$pCOHP plot for the closest carbon and oxygen bond as a function of bonding energy. The Fermi energy is set at zero. The negative and positive values correspond to the antibonding and bonding contributions, respectively. The COHP is found to be very small to claim any kind of covalent bonding.}
\end{figure}

To understand further the bonding nature between the C and O atom, we have carried out Crystal Orbital Hamiltonian Population (COHP) analysis for a pair of closest C and O atom as shown in Fig.~\ref{cohp-tio2}. The magnitude of COHP plot is negligible across the energy window suggesting absence of covalent bonding. 

\providecommand{\latin}[1]{#1}
\makeatletter
\providecommand{\doi}
  {\begingroup\let\do\@makeother\dospecials
  \catcode`\{=1 \catcode`\}=2 \doi@aux}
\providecommand{\doi@aux}[1]{\endgroup\texttt{#1}}
\makeatother
\providecommand*\mcitethebibliography{\thebibliography}
\csname @ifundefined\endcsname{endmcitethebibliography}
  {\let\endmcitethebibliography\endthebibliography}{}

\end{document}